\documentclass{jpp}
\usepackage{graphicx}
\usepackage{epstopdf, epsfig}
\usepackage{hyperref}
\usepackage{natbib}
\usepackage{amsmath}
\usepackage{soul}
\usepackage{xcolor}

\usepackage{color}

\shorttitle{Development of Tearing Instability in a Forming Current Sheet}
\shortauthor{E. A. Tolman, N. F. Loureiro and D. A. Uzdensky}

\title{Development of Tearing Instability in a Current Sheet Forming by Sheared Incompressible Flow}

\author{Elizabeth A. Tolman\aff{1}
  \corresp{\email{etolman@mit.edu}},
 Nuno F. Loureiro\aff{1}
 \and Dmitri A. Uzdensky\aff{2,3}}

\affiliation{\aff{1}Plasma Science and Fusion Center, Massachus
etts Institute of Technology, Cambridge, Massachusetts 02139, USA
\aff{2}Center for Integrated Plasma Studies, University of Colorado, Boulder, Colorado 80309, USA
\aff{3}Institute for Advanced Study, Princeton, NJ 08540, USA}
\begin{document}


\maketitle

\begin{abstract}
Sweet-Parker current sheets in high Lundquist number plasmas are unstable to tearing, suggesting they will not form in physical systems. Understanding magnetic reconnection thus requires study of the stability of a current sheet as it forms. Formation can occur due to sheared, sub-Alfv\'enic incompressible flows which narrow the sheet. Standard tearing theory \citep{furth1963finite,coppi1976resistive,rutherford1973nonlinear} is not immediately applicable to such forming sheets for two reasons: first, because the flow introduces terms not present in the standard calculation; second, because the changing equilibrium introduces time dependence to terms which are constant in the standard calculation, complicating the formulation of an eigenvalue problem.  This paper adapts standard tearing mode analysis to confront these challenges.  In an initial phase when any perturbations are primarily governed by ideal MHD, a coordinate transformation reveals that the flow compresses and stretches perturbations. A multiple scale formulation describes how linear tearing mode theory \citep{furth1963finite,coppi1976resistive} can be applied to an equilibrium changing under flow, showing that the flow affects the separable exponential growth only implicitly, by making the standard scalings time-dependent. In the nonlinear Rutherford stage, the coordinate transformation shows that standard theory can be adapted by adding to the stationary rates time dependence and an additional term due to the strengthening equilibrium magnetic field. Overall, this understanding supports the use of flow-free scalings with slight modifications to study tearing in a forming sheet.
\end{abstract}

\section{Introduction}
Magnetic reconnection is a widely-studied plasma process that rearranges the magnetic topology in a plasma, releasing magnetic energy \citep{biskamp2005magnetic,zweibel2009magnetic,yamada2010magnetic,loureiro2015magnetic}. It is thought to be important to a wide variety of astrophysical and laboratory phenomena, including solar flares \citep{ogawara1994magnetic} and tokamak disruptions \citep{hender2007mhd}. Resistive MHD treatments of magnetic reconnection have typically studied the process in the context of highly regular, intense Sweet-Parker current sheets \citep{parker1957sweet,sweet}, but recent work has shown that such current sheets at high Lundquist numbers are strongly unstable  \citep{loureiro2007instability, samtaney2009formation,bhattacharjee2009fast,loureiro2013plasmoid,loureiro2015magnetic}, suggesting they may never form in the first place \citep{loureiro2007instability,pucci,vekstein2015nonlinear,PhysRevLett.116.105003,comisso2016general}. A better understanding of magnetic reconnection thus requires a model for how this process occurs in a current sheet that develops from a non-reconnecting, stable plasma equilibrium. One potential scenario for such formation is that a stable region is rearranged and compressed by a driving plasma flow. As the plasma starts to form a current sheet, it will gradually become unstable to tearing modes. These tearing modes will evolve in a current sheet that continues to be compressed and stretched by external flows. This could for instance occur during the development of a solar flare \citep{low1988spontaneous} or in substorms in the magnetotail \citep{baker1996neutral}.

\citet{PhysRevLett.116.105003} take a first step in understanding the behavior of such systems by applying growth rates developed for flow-free, stationary magnetic configurations to evolving current sheets, such that the rates change with time in a manner determined by their dependence on current sheet aspect ratio and length. Briefly, in a forming current sheet of length $L\left(t\right)$ and thickness $a\left(t\right)$, where  $L\left(t\right) / a \left(t \right)$ increases with time, one can identify each linear tearing mode by the number of periods of perturbation (magnetic islands) that fit inside the sheet, $N$, so that $k\left( t \right) \sim N / L\left(t \right) $. Inserting this expression for $k\left(t \right)$ into traditional tearing mode growth rates, derived in a static magnetic configuration without flow, reveals the growth rate dependence on sheet aspect ratio $L\left(t \right)/ a\left(t \right)$. 
When a mode is in the small $\Delta ' a\left( t \right)$ (FKR) regime \citep{furth1963finite}, with $\Delta'$ the tearing mode stability index, the linear growth rate is assumed to be given by the classic FKR expression
\begin{equation} \label{eq:oldfkr}
\gamma_{\rm FKR}\left( t \right) \simeq  \tau_A \left( t\right)^{-1} S_a\left(t \right)^{-3/5} N^{-2/5} \left[L(t)/a(t)\right]^{2/5},
\end{equation}
where $\tau_A\left(t \right)$ is Alfv\'en time, $\tau_{A} \left(t \right)\equiv a \left( t \right)/ v_{A}\left(t \right)$, with Alfv\'{e}n speed $v_{A}\left( t \right) \equiv B_{y}\left( t\right)/ \sqrt{4 \pi \rho_0}$ defined in terms of background magnetic field strength $B_{y}\left(t \right)$ and the plasma density $\rho_0$. Here, $S_a\left(t \right)$ is the Lundquist number, $S_a\left( t \right) \equiv v_{A} \left(t \right) a\left(t \right) / \eta$, with $\eta$ the plasma magnetic diffusivity.
The equivalent dependence in the large $\Delta' a\left( t \right)$ (Coppi) regime \citep{coppi1976resistive} is 
\begin{equation}
\label{eq:oldcoppi}
\gamma_{\rm Coppi} \left( t \right) \simeq \tau_A\left(t \right)^{-1} S_a\left(t \right)^{-1/3}N^{2/3} \left[L(t)/a(t)\right]^{-2/3} .
\end{equation}
As the external flow increases the aspect ratio, both growth rates change, and the growth rate and wavenumber of the fastest-growing mode change accordingly. \citet{PhysRevLett.116.105003} analyze this behavior in detail to determine which mode, $N_{\rm max}$, is the fastest growing around the time $t_{\rm cr}$ when the linear tearing mode growth rate overwhelms the current sheet formation rate. They then likewise study the nonlinear stage by inserting aspect ratio dependence into flow-free equations, finding that the first mode to overwhelm the current sheet formation rate is also the first to disrupt the current sheet. 

Though this analysis creates a framework through which the tearing instability in a changing current sheet may be understood, it leaves open the question of how flow and a changing equilibrium affect the derivation of tearing mode growth rates themselves, and whether growth rates derived in stationary configurations can be applied in this way to evolving sheets. The goal of this paper is to more rigorously examine the stability of a current sheet as it forms in order to understand how flow and the resulting change in the equilibrium influence the growth of perturbations, and thereby to justify and suggest modifications to the application of flow-free methods to forming current sheets. This work supports studies of forming current sheets beyond  \citep{PhysRevLett.116.105003} as well. Some recent works \citep{comisso2016general,comisso2017plasmoid,huang2017plasmoid} have considered the onset of reconnection in the context of a principle of least time, which determines the mode that emerges first from the linear phase by minimizing the time taken to traverse the linear phase. Another paper \citep{pucci} computes the growth rate of the plasmoid instability as a function of the current sheet aspect ratio, and then assumes that such a current sheet is forming at the Alfv\'{e}nic rate in order to determine the critical aspect ratio the sheet can reach before it becomes intrinsically unstable. Like \citet{PhysRevLett.116.105003}, these studies use standard tearing mode scalings without rigorously considering how flows and a changing equilibrium might affect the scalings. 

Various papers have studied the effect of certain types of flow on tearing stability. \citet{bulanov1978stabilizing}, \citet{biskamp1986magnetic}, and \citet{bulanov2016magnetic} consider the effect of a compressible flow whose only component is in the direction of the reconnecting field (i.e., along the outflow direction) on tearing growth rate and find that such a flow is stabilizing if its rate exceeds a fraction of the flow-free tearing growth rate.  \citet{chen1990resistive} consider an incompressible flow with components in the direction of the reconnecting field and perpendicular to both the direction of the reconnecting field and the direction in which its strength varies; they find the flow shear to be stabilizing when it is larger than the magnetic field shear at the center of the sheet.
However, these flows do not include a component into the sheet and do not cause any change in the current sheet configuration, making them essentially different from those considered here.

The present paper uses the resistive reduced magnetohydrodynamic (resistive RMHD) framework to consider the effect of incompressible flows driving current sheet formation on the tearing mode. These flows include components into and out of the sheet and self-consistently cause the narrowing and lengthening of the current sheet. The paper proceeds as follows. Section~\S\ref{sec:rmhd} introduces the reduced magnetohydrodynamic (RMHD) framework used to study the tearing mode, and presents an equilibrium for a forming current sheet in this framework;~\S\ref{sec:fourier} discusses a procedure for Fourier analyzing this sheet along its length using a sheared wave vector. The ideal MHD evolution of tearing-type perturbations in this current sheet is studied in Section~\S\ref{sec:ideal} to gain insight into the period before significant tearing occurs. A straightforward coordinate transformation in this case reveals that current sheet formation compresses and stretches the perturbations which would occur in the equivalent static current sheet.  Section~\S\ref{sec:linear} considers linear tearing mode behavior. The ideal MHD coordinate transformation is not sufficient to understand linear tearing dynamics, so~\S\ref{sec:linear} formulates the  perturbed resistive equations in terms of two scales, one that governs the fast tearing behavior, and another that governs the slow equilibrium evolution.  This method, in the spirit of WKB analysis, shows that in the limit of a fast tearing growth rate, the eigenfunctions of the tearing mode grow slowly due to the flow, in addition to undergoing familiar exponential growth at a rate given by standard tearing mode scalings with time dependence added. Section~\S\ref{sec:nonlinear} considers the nonlinear development in the presence of these flows. Here, the same coordinate transformation used in the ideal stage shows that standard tearing mode theory can be adapted to the changing case by adding to the stationary rates time dependence and an additional term due to the shrinking of island width in a strengthening magnetic field. The conclusion places the findings in each stage in the context of the full evolution of the current sheet, including a discussion of the time period around $t_{cr}$ when neither the analysis in~\S\ref{sec:ideal} nor that in~\S\ref{sec:linear} is fully valid. Finally, the conclusion discusses how this analysis influences and supports the techniques used in~\citet{PhysRevLett.116.105003} and works by other authors, and suggests areas for future work.

\section{RMHD description of a forming current sheet}
\label{sec:rmhd}

We study an incompressible plasma with a strong and uniform $z$-directed guide magnetic field, such that the behavior of the $x$-$y$ plane flow and magnetic field can be modeled by the resistive RMHD equations \citep{strauss1976nonlinear, kadomtsev1973nonlinear, schekochihin2009astrophysical}.  These read, in cgs units,
\begin{equation}
\rho_0 \left(\partial_t \nabla_{\perp}^2 \phi + \left \{ \phi, \nabla_{\perp}^2 \phi  \right\} \right)= \frac{1}{c}\left\{ \psi, \frac{c}{4\pi}\nabla_{\perp}^2 \psi \right\}
\end{equation}
and
\begin{equation}
\partial_t \psi + \left \{ \phi, \psi\right\} = \eta \nabla_{\perp}^2 \psi.
\end{equation}
Here, $\phi$ is the stream function of the plasma fluid velocity in the $x$-$y$ plane, $\mathbf{u}$, such that $\mathbf{u} = \left(- \partial_y \phi, \partial_x \phi \right)$, and $\psi$ is the flux function of the in plane magnetic field $\mathbf{B}$, so that $\mathbf{B} = \left(-\partial_y \psi, \partial_x \psi \right)$ (in addition, an asymptotically strong magnetic field in the $\hat{z}$ direction is assumed to be present). The Poisson bracket of two fields $P,Q$ is $\left\{P, Q \right\}= \partial_x P \partial_y Q- \partial_y P\partial_x Q$; the plasma mass density is $\rho_0$.  

We normalize the RMHD equations according to $\psi \rightarrow  \psi/ \sqrt{4 \pi \rho_0}$, $\phi \rightarrow \phi$, $t \rightarrow t $, $x \rightarrow x$, $y \rightarrow y$, and $\eta \rightarrow \eta$.   This non-standard normalization does not include any characteristic length scales or magnetic field strengths to avoid confusion due to changes in these quantities as the current sheet evolves. With these normalizations, the RMHD equations lose their explicit mass density dependence to read:
\begin{equation}
\label{eq:rmhd1}
\partial_t \nabla_{\perp}^2 \phi + \left \{ \phi, \nabla_{\perp}^2 \phi  \right\} = \left\{ \psi,\nabla_{\perp}^2 \psi \right\},
\end{equation}
and 
\begin{equation}
\label{eq:rmhd2}
\partial_t \psi + \left \{ \phi, \psi\right\} = \eta \nabla_{\perp}^2 \psi.
\end{equation}
\begin{figure}
\centering
\includegraphics[width=.6\columnwidth]{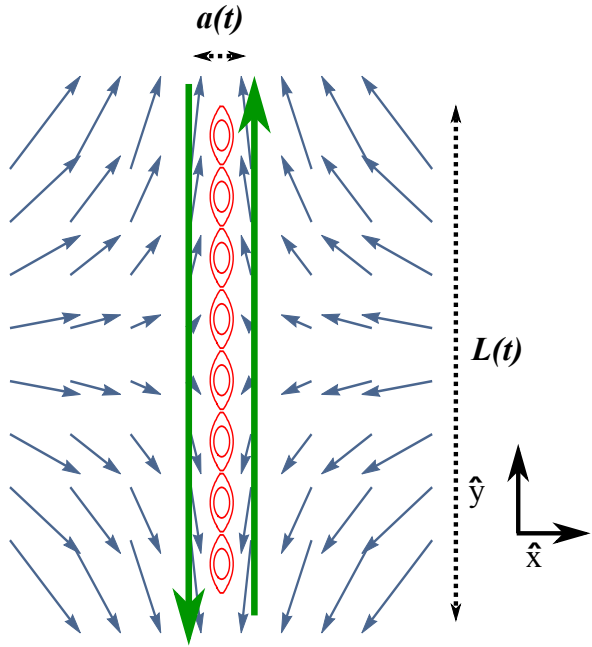}
\caption{Schematic configuration of forming current sheet, which is characterized by a width $a\left(t \right)$ and a length $L\left(t \right)$. Thick green lines represent the equilibrium magnetic field, the blue vector field the flow, and the red magnetic island chain the tearing perturbation.}
\label{fig:sheet}
\end{figure}

We will employ these equations to study an ideal MHD current sheet that forms as a result of plasma flows. This serves as the equilibrium for the paper's tearing stability analysis. The sheet is depicted schematically in Figure~\ref{fig:sheet}. While many specific mathematical descriptions of such a sheet may exist, a clear, analytic tearing stability analysis makes desirable two characteristics: first, that the sheet's magnetic field have no $y$ dependence, in order to allow Fourier analysis in this direction; second, that the sheet's flow pattern be simple enough to allow for the use of coordinate transformations and sheared wave vectors in the tearing calculation. A model that meets these goals is an extension of the one-dimensional Chapman-Kendall \citep{chapman1963liquid} sheet:
\begin{equation}
\label{eq:fluxex}
\psi_0 \left( x, t \right) = B_{0y} a_0 f \left[\frac{x}{a \left( t \right)}\right],
\end{equation}                 
where the current sheet width $a\left(t \right)$ narrows with time according to 
\begin{equation}
\label{eq:widtha}
a\left(t \right)= a_0 e^{-\int_0^t \Gamma_0 \left(s \right)d s},
\end{equation}
with $a_0$ being the initial ($t=0$) width and $\Gamma_0 (t)$ a time-dependent shear parameter.
The magnetic field given by this flux function is
\begin{equation}
\mathbf{B}\left( x,t\right) =B_{0y}e^{\int_0^t \Gamma_0 \left(s \right)d s} f' \left[ \frac{x}{a \left( t \right)}\right] \hat{y} = B_y \left(t \right) f' \left[\frac{x}{a \left( t \right)}\right] \hat{y},
\end{equation}
with
\begin{equation} 
\label{eq:strengthb}
B_y \left(t \right) \equiv B_{0y}e^{\int_0^t \Gamma_0 \left(s \right)d s}
\end{equation}
the characteristic, time dependent, strength of the sheet's magnetic field.
The corresponding stream function is given by                                                                                                                                                                                                                                                                                                                                                                                                                                                                                                                                                                                                                                                                                                                                                                                                                                                                                                                                                                                                                                                                                                                                                                                                                                                              
\begin{equation}
\label{eq:streamex}
\phi_0 \left(x,y,t \right) =\Gamma_0 \left(t \right) xy.
\end{equation}
In our study of the ideal MHD behavior (Section~\S\ref{sec:ideal}), our analysis is limited to the specific case of constant $\Gamma_0 \left( t \right) = \Gamma_0$, while in our study of linear and nonlinear tearing (Sections~\S\ref{sec:linear} and~\S\ref{sec:nonlinear}), our analysis applies for slowly-varying $\Gamma_0\left( t\right)$ as well.
Functions (\ref{eq:fluxex})  and (\ref{eq:streamex}) together obey (\ref{eq:rmhd1}) and (\ref{eq:rmhd2}) in the ideal limit $\eta = 0$ for any magnetic field shape $f$. 
The stream function $\phi_0\left(x,y, t \right)$ is the same used in the Chapman-Kendall model and represents a general expansion of flow about a stagnation point. For functions $f$ yielding magnetic fields which are locally linear about $x=0$, the first term in the expansion of (\ref{eq:fluxex}) is the Chapman-Kendall flux function, $\psi_0 \left( x, t \right) = B_{0y} a_0 x^2/\left[ 2 a \left(t \right)^2\right]$. 
At times, our calculation will require a specific form for the magnetic field; in these cases, we will often take the $f$ that yields a Harris sheet~\citep{harris1962plasma}, such that 
\begin{equation}
\label{eq:harris}
\psi_0 \left( x, t \right) = B_{0y} a_0 \log \left( \cosh\left[\frac{x}{a\left(t \right)} \right]\right),
\end{equation}
and 
\begin{equation}
\mathbf{B}\left( x,t\right) =B_y\left(t \right)\tanh{\left[\frac{x}{a\left(t \right)}\right]} \hat{y}.
\end{equation}

We have constrained the equilibrium magnetic field to only have a component in the $\hat{y}$ direction, the magnitude of which is independent of $y$, to enable Fourier analysis in $y$. Thus, the length of our current sheet is not mathematically defined by the expressions given so far. However, the current sheet is stretched by a flow along its length, $u_y = \Gamma_0\left(t\right)y$, which grows with distance along the sheet $y$ in the same sense that the flow into the sheet, which causes current sheet compression, $u_x  = -\Gamma_0 \left( t \right)x$, decreases with decreasing $x$.  We may thus define an effective length 
 \begin{equation}
 \label{eq:loft}
 L  \left( t \right)\sim L_0 \frac{a_0}{a \left(t \right)} = L_0e^{\int_0^t \Gamma_0 \left(s \right)ds},
 \end{equation}
 which grows as the width of the current sheet shrinks.  This definition leaves the product $a\left(t \right)L\left(t \right)$ constant, consistent with incompressibility.  Another way to think about this length is to consider placing two markers in the current sheet that move with the flow.  If the markers are initially separated in $y$ by a distance $L_0$, their future separation is given by (\ref{eq:loft}). This exponential behavior of the current sheet width and length agrees with trends seen in simulations \citep{sulem1985exponential,friedel1997adaptive,biskamp2005magnetic}.
 
Thus, current sheet formation in this scenario proceeds through a combination of thinning ($a$ decreasing in time, as given by (\ref{eq:widtha})), stretching ($L$ increasing with time, as given by (\ref{eq:loft})), and magnetic field strengthening ($B_y$ increasing with time, given by (\ref{eq:strengthb})). The exponential growth in the magnetic field strength is consistent with the exponential growth and collapse of the spatial scales. All $x$-scales and separations decrease in time exponentially; this includes the $x$-separation between two given magnetic fields lines frozen into the fluid, and this is the reason the magnetic field strength is increasing.
 
\section{Fourier analysis of a forming current sheet}
\label{sec:fourier}
The aim of this paper is to study the behavior of a small perturbation to the forming current sheet described by (\ref{eq:fluxex}) and (\ref{eq:streamex}), such that $\psi\left(x,y,t\right) = \psi_{0}\left( x,t \right) + \delta \psi \left(x,y,t\right)$ and $\phi \left(x,y,t\right) = \phi_{0}\left(x,y,t\right) + \delta \phi\left(x,y,t\right)$.
Evaluating (\ref{eq:rmhd1}) and (\ref{eq:rmhd2}) for the behavior of the perturbation in linear order yields

\begin{multline}
\label{eq:perturb1}
 \partial_t \left(\partial_x^2 + \partial_y^2 \right) \delta \phi +  \left( \partial_x \phi_0  \right) \, \left[ \partial_y \left( \partial_x^2 + \partial_y^2\right)\delta\phi \right]
-\left( \partial_y \phi_0 \right) \, \left[ \partial_x \left( \partial_x^2 + \partial_y^2\right)\delta \phi \right]
\\ = \left(\partial_x \psi_0 \right) \, \left[\partial_y \left(\partial_x^2 + \partial_y^2 \right) \delta \psi \right] - \left(\partial_y \delta \psi \right) \, \left(\partial_x^3 \psi_0\right)
\end{multline}
and 
\begin{equation}
\label{eq:perturb2}
\partial_t \delta\psi  + \left( \partial_x \phi_0 \right) \, \left( \partial_y \delta  \psi \right) - \left(\partial_y \delta \phi \right) \, \left(\partial_x \psi_0\right) - \left( \partial_y \phi_0 \right) \left( \partial_x \delta \psi \right)= 
\eta \left(\partial_x^2 + \partial_y^2 \right) \delta \psi.
\end{equation}
We Fourier analyze these perturbations along the $\hat{y}$ direction, such that
 $\delta \phi = \phi_1 \left(x,t \right) \exp \left[ i k\left(t\right)y\right]$ and $\delta \psi = \psi_1 \left(x,t \right) \exp \left[ i k\left(t\right)y\right]$. 
Because the current sheet is evolving, we take a wave vector $k\left(t \right)$ that also evolves.  The postulate for $k\left(t \right)$ is guided by the intuition that the number of islands $N$ characterizing a given mode should remain constant as the current sheet stretches, such that
\begin{equation}
\label{eq:wavevec}
k \left( t \right) = k_0 e^{- \int_0^t \Gamma_0 \left(s \right)ds}\sim N/L\left(t \right).
\end{equation}
This postulate also allows the cancellation of the explicit $y$-dependence in terms containing $\partial_x \phi_0$ in (\ref{eq:perturb1}) and (\ref{eq:perturb2}).  With this choice, a critical simplification in enabling the analysis that follows, (\ref{eq:perturb1}) and (\ref{eq:perturb2}) using (\ref{eq:streamex}) reduce to 
\begin{multline}
\label{eq:one}
\left[ \partial_x^2 -k^2\left(t \right) \right] \partial_t \phi_1 - \Gamma_0\left(t \right) x \partial_x \left[ \partial_x^2 
- k^2\left(t \right)\right]\phi_1 + 2 \Gamma_0\left(t \right) k^2\left(t \right) \phi_1 
 \\
=
\partial_x \psi_0ik\left(t\right) \left[\partial_x^2 - k^2\left(t \right) \right]  \psi_1 
-\partial_x^3 \psi_0 i k\left(t \right) \psi_1
\end{multline}
and 
\begin{equation}
\label{eq:ohm}
\partial_t \psi_1 - \Gamma_0\left(t \right) x \partial_x \psi_1 - \partial_x \psi_0 ik\left( t \right) \phi_1 
= \eta \left[\partial_x^2 - k^2\left( t \right) \right] \psi_1.
\end{equation}
Sections~\S\ref{sec:ideal} and~\S\ref{sec:linear} are concerned with solving these equations.

\section{Effect of flow on perturbations in ideal MHD}
\label{sec:ideal}

The main aim of this paper is to analyze the resistive tearing dynamics of a current sheet as it forms due to flow. Before considering this topic, however, it is worthwhile to examine the \textit{ideal} behavior of small perturbations that occur in a forming current sheet.  These are changes due to the effect of the flow alone and not due to resistive tearing, which remain when $\eta$ is set to zero in (\ref{eq:ohm}).  This behavior describes the system when the gradients in the forming current sheet are not yet sufficient to drive a tearing instability whose growth rate overwhelms the flow shear rate, so any tearing that may occur is a small perturbation to the dominant flow-driven behavior.  Our focus is on tearing-like modes: those that are even with respect to the midplane $x= 0$ and that are associated with non-zero perturbed flux at $x = 0$. We find that these perturbations are compressed and stretched by the flow. However, we also note in passing a way in which this analysis may be adapted to treat other types of modes.

The ideal MHD behavior of our current sheet may be studied by setting $\eta = 0$ in (\ref{eq:ohm}). In addition, in this section, we consider the special case of constant flow rate,
\begin{equation}
\Gamma_0\left(t \right)= \Gamma_0.
\end{equation}
We now perform a transformation of coordinates in Eqs. (\ref{eq:one}) and (\ref{eq:ohm}) such that
\begin{equation}
\label{eq:transformation}
\left(x,t \right)\rightarrow \left(x' = x e^{\Gamma_0 t}, t' = t \right),
\end{equation}
which moves the system to coordinates which are compressed with the flow. The transformation gives
\begin{multline}
\label{eq:fullone}
\partial_{t'} \left(e^{2\Gamma_0 t' }\partial_{x'}^2  -  e^{-2\Gamma_0 t'} k_0^2\right) \phi_1
 \\ 
 =  e^{2 \Gamma_0 t'}\partial_{x'} \psi_0 ik_0 \partial_{x'}^2\psi_1 
 -  e^{-2\Gamma_0 t'}\partial_{x'} \psi_0 ik_0\left( k_0^2 \right) \psi_1
 - e^{2\Gamma_0 t'}\partial_{x'}^3 \psi_0 i k_0 \psi_1,
\end{multline}
and 
\begin{equation}
\label{eq:fullohm}
\partial_{t'} \psi_1  - \partial_{x'} \psi_0 ik_0\phi_1 
=0.
\end{equation}
Solving (\ref{eq:fullohm}) for $\phi_1$ and substituting into (\ref{eq:fullone}) then yields a differential equation for $\psi_1$.
For each mode $k_0$, there will come a time at which $e^{2\Gamma_0 t'} \partial_{x'}^2 \gg e^{-2 \Gamma_0 t'} k_0^2$.  After this time, which we discuss at the end of this section, the terms proportional to $k_0^2$ may be dropped from (\ref{eq:fullone}). 
When this is possible, the differential equation for $\psi_1$ may be Fourier analyzed in time $t'$, such that $\psi_1 = \Psi_1\left(x' \right) e^{-i\omega t'}$. This gives the following equation relating $\omega$ and the derivatives of $\Psi_1$, where we have plugged in (\ref{eq:fluxex}) for the equilibrium flux function:
\begin{equation}
\label{eq:transformed}
\frac{\omega^2 + 2 i \omega \Gamma_0}{k_0^2 B_{0y}^2}\partial_{x'}^2 \left[\frac{\Psi_1}{ f'\left(\frac{x'}{a_0}\right) }\right]
-\left[ f' \left(\frac{x'}{a_0} \right)\partial_{x'}^2 \Psi_1 - \frac{1}{a_0^2}f'''\left(\frac{x'}{a_0}\right)\Psi_1\right] = 0.
\end{equation}
The solutions to this differential equation will depend on the specific form of $f$. However, the equation alone shows the general effect of flow on perturbations that exist in the current sheet. 
Equation~(\ref{eq:transformed}) demonstrates that under the transformation~(\ref{eq:transformation}), the behavior of the perturbations appears identical to perturbations in a flow-free system (consider setting $\Gamma_0 \rightarrow 0$ in~(\ref{eq:transformed})) except for the redefinition
\begin{equation}
\label{eq:relat}
\omega_{0}^2 \rightarrow \omega^2 + 2 i \omega \Gamma_0, 
\end{equation}
where $\omega$ is the frequency of the perturbation in the current sheet evolving due to flow and $\omega_{0}$ is the corresponding frequency in a static magnetic configuration.  Thus, for every mode that exists in a static magnetic configuration (equation~(\ref{eq:fluxex}) with $\Gamma_0 = 0$) of frequency $\omega_0$, we can find a corresponding mode in the evolving current sheet. The mode in the developing current sheet has the same structural form as the static mode, but is expressed as a function of the transformed coordinates and has a modified frequency $\omega$.  This modified frequency is  found by solving~(\ref{eq:relat}) for $\omega$ as a function of the corresponding $\omega_{0}$ and $\Gamma_0$, so that
\begin{equation}
\label{eq:freq}
\omega =  - i \Gamma_0 + \sqrt{\omega_0^2 - \Gamma_0^2},
\end{equation}
which reduces, as expected, to $\omega = \omega_0$ in the case of no flow.\footnote{We will presently focus on modes for which $\omega = 0$ and $\omega_0 = 0$. However, we have chosen to leave this equation in its full form to suggest how the analysis can be modified for other types of modes.
Plugging~(\ref{eq:freq}) into $\psi_1 = \Psi_1 \left( x'\right) e^{-i \omega t'}$ and transforming back to unprimed coordinates gives
$\psi_1 \left(x,t\right) = \Psi_{1,0} \left( x e^{\Gamma_0 t}\right)e^{-\Gamma_0 t} e^{- i\left(\sqrt{\omega_{0}^2 - \Gamma_0^2}\right)t}$,
where $\omega_{0}$ and $\Psi_{1,0}$ are the frequencies and eigenfunctions of the stationary magnetic configuration corresponding to the evolving current sheet.  This indicates that in the general case, modes that exist in an evolving system are related to those that exist in a flow-free configuration by a self-similar collapse and a change in frequency. Like any inhomogeneous magnetic field, a current sheet magnetic field may be subject to Alfv\'{e}nic modes with eigenfunctions sharply localized at points where the mode frequency matches the local Alfv\'{e}n frequency~\citep{goedbloed2004principles}, and our analysis can be applied to these modes as well, to show how their structure and frequency are altered.}

We would like to consider tearing-like modes, for which $\Psi_1\left(0 \right) \neq 0$.  These modes are more straightforward than other modes which may exist in the sheet.  Consider evaluating equation~(\ref{eq:fullohm}) at $x' = 0$.  Since the magnetic field passes through 0, and hence  $\partial_{x'} \psi_0|_{x' = 0} = 0$, this evaluation immediately reveals that $\partial_{t'} \psi_1|_{x' = 0}  = 0$. In ideal MHD, magnetic flux is conserved, and since the center of the sheet has no equilibrium velocity or magnetic field, the value of the flux there must be constant. Separating the perturbation according to $\psi_1 =\Psi_1 \left( x'\right) e^{-i \omega t'}$ then gives two possibilities: either $\Psi_1\left(0\right) = 0$ or $\omega = 0$.  While in general modes in the sheet may obey either condition, tearing-like modes must have $\Psi_1\left(0\right) \neq 0$ and hence $\omega = 0$. Thus, for these modes,~(\ref{eq:transformed}) reduces to:
\begin{equation}
\label{eq:idealtearing}
 f' \left(\frac{x'}{a_0} \right)\partial_{x'}^2 \Psi_1 - \frac{1}{a_0^2}f'''\left(\frac{x'}{a_0}\right)\Psi_1 = 0.
\end{equation}
This equation, stated in the primed coordinates, is unmodified from its flow-free equivalent, so we see that the effect of the flow on these modes is simply to compress them as represented by the primed coordinates. (The sheared wave vector taken in the Fourier analysis of Section~\S\ref{sec:fourier} will also stretch the islands.) Note that it also follows from~(\ref{eq:fullohm}) and $\omega =0$ that $\phi_1=0$; there is no flow, perturbed or equilibrium, in the primed coordinates. 

\begin{figure}
\centering
\includegraphics[width=.5\columnwidth]{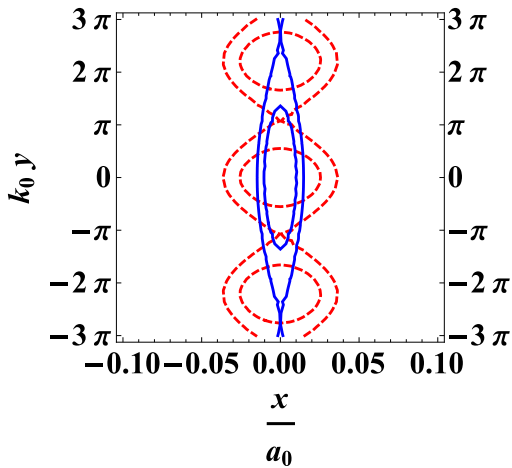}
\caption{Contour lines of $\psi_0 + \psi_1$ for a tearing-like mode undergoing ideal evolution. The perturbed flux, $\psi_1$, is given by~(\ref{eq:coseig}), while $\psi_0$ is given by $\psi_0 = - B_{0y} a_0 \cos{\left( \left[\pi/2 \right] \left[ x/  a\left(t\right)\right]\right)}$, $a \left(t \right) = a_0 e^{-\Gamma_0t}$. The configuration is shown for $\Gamma_0 t = 0.1$ (dashed red line) and $\Gamma_0 t = 1$ (solid blue line). The strength of the perturbation relative to the equilibrium magnetic fields is given by $\psi_1\left(0 \right)/B_{0y}= 0.001$. The islands are compressed in $x$ and lengthened in $y$ by the flow, just as the sheet width and wave vector are; indeed, if the islands were instead plotted in terms of $k\left(t \right) y $ and $x/ a\left(t \right)$, the islands would appear to be constant. }
\label{fig:idealislands}
\end{figure}
As an example, we can select the flux function with $f \left[x/a\left(t\right) \right] = -\cos{\left( \left[\pi /2 \right] \left[x/a\left(t\right)\right]\right)}$, which gives the magnetic field $\mathbf{B} =  \hat{y} B_{0y}e^{\Gamma_0t}\pi  \sin{\left( \left[\pi /2 \right] \left[x/a\left(t\right)\right]\right)}  /2 $, which is locally linear about $x = 0$.
This configuration allows a simple solution (unlike the Harris equilibrium~(\ref{eq:harris})).  We solve~(\ref{eq:idealtearing}) for $\Psi_1$, constraining our solution to be even in $x'$.  Stating the perturbed flux function in terms of the unprimed coordinates then gives
\begin{equation} 
\label{eq:coseig}
\psi_1 \left(x,t\right) = \psi_1\left(0 \right)\cos{\left(\frac{\pi x}{2a_0 e^{-\Gamma_0 t}}\right)},
\end{equation}
with $\psi_1\left(0 \right)$ a constant. We see here the expected self-similar compression; the mode will also be stretched by the sheared wave vector. This behavior is illustrated in Figure~\ref{fig:idealislands}.

The analysis in this section is only valid once $e^{2\Gamma_0 t'} \partial_{x'}^2 \gg e^{-2 \Gamma_0 t'} k_0^2$.  Taking a typical value $\partial_{x'}^2 \sim 1/a_0^2$ (as is true in~(\ref{eq:coseig})), we see that this condition is just that $k\left(t \right) a \left( t \right)\ll 1$; in other words, for this analysis to be valid, the mode's wavelength $\lambda\left(t \right) = 2 \pi / k \left(t \right)$ must be much longer than the sheet thickness at the same time, $a\left(t \right)$. Since a mode's wavelength grows with time while the sheet thickness shrinks with time, this  condition will become better satisfied for any mode as time increases.  On the other hand, as time increases, the resistive term in~(\ref{eq:ohm})  will also become more important, and the sheet's tendency to tear will increase.  Our ideal analysis will only be useful for modes for which the condition $k\left(t \right) a \left( t \right)\ll 1$ is satisfied before resistivity becomes important. We can estimate that resistivity becomes important at the time $t_{\rm cr}$ when the flow-free tearing rate of the fastest growing mode becomes equal to the flow shear rate $\Gamma_0$.  The growth rate of the fastest-growing mode can be found by balancing the expressions in~(\ref{eq:oldfkr}) and~(\ref{eq:oldcoppi}), which have opposite scaling with $k$, to give $\gamma_{0,\rm max} = \eta^{1/2}B_y^{1/2}/a^{3/2}$. This rate becomes equal to $\Gamma_0$ when  $t = t_{cr} = \log{\left[\left(\Gamma_0 a_0^{3/2}\right)/\left( \eta^{1/2} B_{0y}^{1/2} \right)\right]}/\left(2 \Gamma_0\right)$. Meanwhile, $k\left(t \right) a \left( t \right) = 1$ when  $t  = \log{\left(k_0^2 a_0^2 \right)}/\left(4 \Gamma_0\right)$. Thus, wave vectors for which the ideal analysis is valid at least for some period of time (that is, for which the $k_0^2$ terms may be dropped prior to resistivity becoming important) are given by $k_0 a_0 \ll \left(\Gamma_0 a_0/B_{0y} \right) \left(B_{0y} a_0/ \eta \right)^{1/2}\equiv k_{0, \rm ideal}a_0$, which for high Lundquist number plasmas is a weak limitation.\footnote{To contextualize its value, the wave vector's magnitude  as the sheet evolves may be compared to that of the fastest growing tearing wave vector, $k_{\rm max}\left(t \right) = S_a\left(t\right)^{-1/4}/a\left(t \right)$ \citep{PhysRevLett.116.105003}: $k_{0, \rm ideal}e^{-\Gamma_0 t}/k_{\rm max} \left(t \right) = e^{-2\Gamma_0 t} \left(\Gamma_0 a_0 /B_{0y}\right)S_{a,0}^{3/4}$ with $S_{a,0}$ the initial value of $S_a\left(t \right)$.} 
 Nonetheless, the analysis does not answer the question of what happens for modes with $k_0 a_0 \gg \left(\Gamma_0 a_0/B_{0y} \right) \left(B_{0y} a_0/ \eta \right)^{1/2}$ or what happens to modes of shorter wavelength before $e^{2\Gamma_0 t'} \partial_{x'}^2 \gg e^{-2 \Gamma_0 t'} k_0^2$.  For these cases, we rely on the intuition that the flow should still stretch and compress tearing-type modes, since it does not present an instability source (see below). However, the form of this stretching and compressing, and the resulting deformation of the tearing mode magnetic islands, will be more complicated than the simple picture in this section.

No modes in the sheet, including those for which the analysis is not valid, will be Kelvin-Helmholtz unstable, for two reasons.  First, the specific flow pattern that we choose,~(\ref{eq:streamex}), lacks the inflection points necessary for Kelvin-Helmholtz instability; second, even if we were to relax our flow model to include patterns with inflection points, for flows that remain sub-Alfv\'enic with respect to~(\ref{eq:strengthb}), no instability would exist \citep{loureiro2013plasmoid}. 

However, the flow will affect perturbations, and the aim of this section has been to describe this effect, showing that they are reshaped ideally by the flow. This is the dominant behavior of any perturbations in the sheet before narrowing $a\left(t \right)$ and sharpening gradients lead to significant tearing.  This tearing stage is treated next.

\section{Linear tearing instability}
\label{sec:linear}
As the width of the current sheet decreases, the importance of the resistive term in the induction equation increases until appreciable reconnection begins to occur.  After this time, reconnection will proceed in a manner modified by flow, first traversing the linear stage of the tearing mode. Previous studies of the effect of flow on the linear tearing mode studied static configurations \citep{bulanov1978stabilizing,paris1983influence,biskamp1986magnetic,chen1990resistive,bulanov2016magnetic}. In these cases, the flow alters the separable exponential growth of the tearing mode; that is, $\psi_1$ may be written as 
\begin{equation}
\label{eq:previous}
\psi_1\left(x,t\right) = \Psi_1\left(x\right)e^{\left[\gamma_0 +  h\left(\Gamma_0\right) \right]t},
\end{equation}
where $\gamma_0$ is the tearing growth rate without flow, and $h$ is some function of flow strength $\Gamma_0$ which either increases or decreases mode growth rate.  

We argue, however, that this formulation is not a proper way of describing the effect of flows that cause change in the equilibrium configuration.  Such flows introduce time dependence into Eqs.~(\ref{eq:one}) and~(\ref{eq:ohm}), such that the perturbation cannot immediately be separated into $x$-dependent and $t$-dependent parts. Applying the coordinate transformation~(\ref{eq:transformation}), as we did when considering ideal MHD behavior, does not resolve this problem, since the transformation leads the resistive term in~(\ref{eq:ohm}) to develop explicit time dependence, even in the case of constant $\Gamma_0$.  Instead, exponential time dependence must be separated out using a multiple scale formulation. The application of this approach reveals that flows add slow time dependence to the standard eigenfunction $\Psi_1\left(x\right)$; however, they do not alter the separable exponential growth.

Let us study the effect of flow on the linear tearing rate by formulating the system in terms of two time scales, one (shorter) which governs the growth of the perturbation and another (longer) which governs the evolution of the current sheet  \citep{bender2013advanced}. When the first of these time scales is much shorter than the second, the ratio of the two scales can be used as a small parameter in which to expand $\psi_1$ and $\phi_1$. The time in current sheet development when this approach is valid, and its relationship to the overall reconnection of the current sheet, are discussed in Section~\S\ref{sec:story}.
We again assume for simplicity $\Gamma_0\left(t \right) = \rm constant$, though the analysis that follows may in principle be generalized to any relatively slowly evolving $\Gamma_0\left( t\right)$ for which $\partial_t \Gamma_0 \left( t \right)/ \Gamma_0 \lesssim \Gamma_0 \left(t \right)$.  
 Let us define:
 \begin{equation}
 \tau \equiv \Gamma_0 t
 \end{equation}
and 
\begin{equation}
\hat{t} \equiv \gamma_0 t,
\end{equation} 
with $\gamma_0$ a value representing the characteristic strength of the tearing growth, which could for instance be the flow-free linear tearing mode growth rate calculated with the values of $k$, $a$, and $B_y$ instantaneously assumed by the mode and the forming current sheet. 
These timescales are related by the small parameter
\begin{equation}
\epsilon \equiv \frac{\tau}{\hat{t}} =\frac{\Gamma_0}{\gamma_0}.
\end{equation}
This quantity will decrease with time as the current sheet narrows and the tendency to tear increases while the flow strength remains constant.  Quantities influenced only by the flow, $k\left(t \right)$ and $\psi_0 \left(x,t\right)$, vary on the slow timescale $\tau$, while the perturbation varies on the fast time scale $\hat{t}$. This separation was not possible in the ideal stage, when the timescale of the perturbation was comparable to the overall sheet evolution timescale. 

In terms of these new timescales,~(\ref{eq:one}) and~(\ref{eq:ohm}) read
 \begin{multline}
\label{eq:onet}
 \left[\partial_x^2 - k^2\left(\tau\right)\right]\gamma_0\partial_{\hat{t}} \phi_1 \left(x,\hat{t}\right)
- \Gamma_0 x \partial_x \left[ \partial_x^2 - k^2\left(\tau\right)\right]\phi_1\left(x,\hat{t} \right)  
+ 2 \Gamma_0  k^2\left( \tau \right) \phi_1 \left(x,\hat{t}\right) \\
= \partial_x\psi_0 \left(x,\tau \right) i k\left(\tau \right)\left[\partial_x^2- k^2\left( \tau \right) \right]\psi_1\left(x, \hat{t}\right) 
- \partial_x^3 \psi_0\left(x, \tau \right) i k\left( \tau \right) \psi_1\left(x,\hat{t} \right)
,
\end{multline}
and 
\begin{equation}
\label{eq:ohmt}
\gamma_0\partial_{\hat{t}} \psi_1\left(x,\hat{t}\right) - \Gamma_0x \partial_x \psi_1\left(x,\hat{t}\right) - \partial_x \psi_0 \left(x,\tau \right) ik \left( \tau \right)\phi_1 \left(x,\hat{t}\right)= \eta \left[ \partial_x^2 - k^2 \left(\tau\right)\right]\psi_1\left(x,\hat{t}\right).
\end{equation}          
Writing these equations in terms of $\tau$ and $\epsilon$ yields 
\begin{multline}
\label{eq:onetau}
\epsilon \left[\partial_x^2 - k^2\left(\tau\right)\right]\gamma_0\partial_{\tau} \phi_1 \left(x,\tau\right)
- \epsilon \gamma_0  x \partial_x \left[ \partial_x^2 - k^2\left(\tau\right)\right]\phi_1\left(x,\tau \right)  
+ 2 \epsilon \gamma_0  k^2\left( \tau \right) \phi_1 \left(x,\tau \right)
 \\
 = \partial_x\psi_0 \left(x,\tau \right) i k\left(\tau \right)\left[\partial_x^2- k^2\left( \tau \right) \right]\psi_1\left(x, \tau \right) 
- \partial_x^3 \psi_0\left(x, \tau \right) i k\left( \tau \right) \psi_1\left(x,\tau \right)
,
\end{multline}
and 
\begin{equation}
\label{eq:ohmtau}
\epsilon \gamma_0\partial_{\tau} \psi_1\left(x,\tau\right) - \epsilon \gamma_0 x \partial_x \psi_1\left(x,\tau\right) - \partial_x \psi_0 \left(x,\tau \right) ik \left( \tau \right)\phi_1 \left(x,\tau\right)
= \eta \left[ \partial_x^2 - k^2 \left(\tau\right)\right]\psi_1\left(x,\tau\right).
\end{equation}

We can then write our solution as an expansion in $\epsilon$, so that
\begin{equation}
\phi_1\left(x,\tau\right) 
= \left[ \Phi_{1,0} \left(x,\tau\right) + \epsilon  \Phi_{1,1} \left(x,\tau\right)  + \ldots + \epsilon^n  \Phi_{1,n} \left(x,\tau\right) \right] \times \exp{\left[\frac{1}{\epsilon} \int_0^\tau \hat{\gamma}\left(s\right) ds \right]},
\end{equation}
and equivalently for $\psi_1\left(x,\tau\right)$. This expansion separates the fast tearing mode growth from the slow evolution of the perturbations, which is in general a non-separable function of space and time.
Note that here $\hat{\gamma}\left(s\right)$ represents $\mathcal{O}\left( 1\right)$ corrections to the growth rate $\gamma_0$, such that the full growth rate in units of $t$ instead of $\hat{t}$ reads $\gamma \left(\tau \right) = \gamma_0 \hat{\gamma}\left(\tau \right)$.  

These expansions may be plugged in to~(\ref{eq:onetau}) and~(\ref{eq:ohmtau}) to yield equations governing the perturbation at each order in $\epsilon$. The $\mathcal{O} \left( \epsilon^0\right)$ equations read:
\begin{multline}
\gamma_0 \hat{\gamma }\left( \tau \right) \left[\partial_x^2 - k^2\left(\tau\right)\right] \Phi_{1,0} \left(x,\tau\right)
 \\
= \partial_x\psi_0 \left(x,\tau \right) i k\left(\tau \right)\left[\partial_x^2- k^2\left( \tau \right) \right]\Psi_{1,0}\left(x, \tau \right) 
- \partial_x^3 \psi_0\left(x, \tau \right) i k\left( \tau \right) \Psi_{1,0}\left(x,\tau \right)
,
\end{multline}
and 
\begin{equation}
\gamma_0 \hat{\gamma} \left( \tau \right)\Psi_{1,0}\left(x,\tau\right) - \partial_x \psi_0 \left(x,\tau \right) ik \left( \tau \right)\Phi_{1,0} \left(x,\tau\right)
\\= \eta \left[ \partial_x^2 - k^2 \left(\tau\right)\right]\Psi_{1,0}\left(x,\tau\right).
\end{equation}       
These equations are identical to the classic flow-free tearing mode eigenvalue equations~\citep{furth1963finite}, but with time-dependent coefficients, growth rates $\gamma\left( \tau \right) = \gamma_0 \hat{\gamma}\left( \tau \right)$, and eigenfunctions $\Psi_{1,0} \left( x, \tau \right)$.   Therefore, the growth rates and eigenfunctions have the same form as their flow-free equivalents, but with $k$, $B_y$, and $a$ modified to be time dependent.  Corrections to the growth rate which are explicitly proportional to $\Gamma_0$ do not appear because terms in the equations governing the perturbation which are explicitly proportional to  $\Gamma_0$ are of the same order as the time dependence of  $k$ and $\partial_x \psi_0$ (which prevents the equations from being completely separable). The tearing growth rate, which is the separable part of the perturbation growth, does not depend explicitly on $\Gamma_0$, in contrast to tearing growth rates in sheets with flow that does not alter the equilibrium.

Therefore, the strongest effect of flows causing current sheet formation on the tearing mode is just to add time dependence to the tearing mode growth rates and eigenfunctions~\citep{furth1963finite,coppi1976resistive}, such that to zeroth order in $\epsilon$
\begin{equation} 
\phi_1\left(x,t\right) = \Phi_{1,0}\left(x,t\right) e^{\int_0^t \gamma\left(s\right)ds},
\end{equation}
and equivalently for $\psi_1$,
where the growth rate can be replaced by the classical expressions

\begin{equation}
\label{eq:modFKR}
\gamma_{ \rm FKR} \left(s \right) = \eta^{3/5} a^{-2}\left(s \right) B_y^{2/5}\left(s \right) k\left(s \right)^{-2/5}/\tau_A\left(s\right),
\end{equation} 
\begin{equation}
\label{eq:modcop}
\gamma_{ \rm Coppi} \left(s\right) = \eta^{1/3} a^{-2/3}\left(s\right) B_y^{2/3}\left(s \right) k^{2/3}\left(s \right)/\tau_A\left(s \right),
\end{equation}
or
\begin{equation}
\label{eq:modmax}
\gamma_{\max} \left(s \right) = S_a^{-1/2} \left(s \right)/\tau_A\left(s\right), 
\end{equation}
depending on whether the mode in question is an FKR mode, a Coppi mode, or the fastest growing mode, found by balancing the Coppi and FKR expressions.
The eigenfunctions read
\begin{equation}
\label{eq:changeig}
\Psi_{1,0} \left(x,t\right) = \Psi_1\left[x; k\left(t \right), a\left(t \right)\right], \, \Phi_{1,0} \left(x,t\right) = \Phi_1\left[x; k\left(t \right), a\left(t \right)\right],
\end{equation}
with $\Psi_1$ and $\Phi_1$ the normal flow-free tearing eigenfunctions, which depend on $k$ and $a$.
While the flow will not affect the separable exponential growth of the perturbation, it will alter how the perturbation changes in time in a non-separable way through its addition of time dependence to the eigenfunctions.
The strength of this correction is given by straightforward differentiation of~(\ref{eq:changeig}):
\begin{equation}
\label{eq:changeeig}
\frac{\partial_t \Psi_{1,0} \left(x,t\right)}{\Psi_{1,0} \left(x,t\right)} = \frac{1}{\Psi_{1,0}}\left( \partial_k \Psi_1 \partial_t k + \partial_a \Psi_1 \partial_t a\right)
\end{equation}
and
\begin{equation}
\label{eq:changeeig2}
 \frac{\partial_t \Phi_{1,0} \left(x,t\right)}{\Phi_{1,0} \left(x,t\right)} = \frac{1}{\Phi_{1,0}}\left( \partial_k \Phi_1 \partial_t k + \partial_a \Phi_1 \partial_t a\right),
\end{equation}
both of which will be proportional to $\Gamma_0$ because $\partial_t k$ and $\partial_t a$ are.

As an example, we can evaluate this effect for the magnetic perturbation in the Harris sheet~(\ref{eq:harris}). Recall that the magnetic eigenfunction in the flow-free case is given by 
\begin{equation}
\left[ \partial_x \psi_0 \left(\partial_x^2 - k^2 \right) - \partial_x^3 \psi_0\right] \Psi_1 = 0,
\end{equation}
with the boundary condition that the eigenfunction go to zero at $x = \pm \infty$ \citep{furth1963finite}. For~(\ref{eq:harris}), the appropriate eigenfunction is 
\begin{equation}
\label{eq:harriseig}
\Psi_1 \left(x; k,a\right) = e^{-k \left| x\right|} \left(1 + \frac{1}{ka}\tanh{\frac{\left|x \right|}{a}} \right).
\end{equation}
\begin{figure}
\centering
\includegraphics[width=.4\columnwidth]{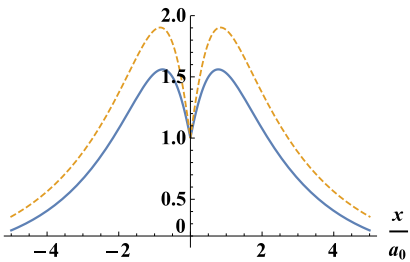}
\caption{Evolution of $\Psi_{1,0}\left(x,t\right)$ for the Harris sheet eigenfunction~(\ref{eq:harriseig}), for $k\left(t\right)$ and $a\left(t\right)$ given by~(\ref{eq:widtha}) and~(\ref{eq:wavevec}) with $\Gamma_0 \left(t \right) = \rm constant$, for $k_0 a_0= 0.5$. The solid blue line is evaluated at $t= 0$, while the orange dotted line is evaluated at $\Gamma_0 t = 0.1$. }
\label{fig:psi1}
\end{figure}From here, we can evaluate that 
\begin{multline}
\frac{\partial_t \Psi_{1,0} \left(x,t\right)}{\Psi_{1,0} \left(x,t\right)} 
=  \Gamma_0 \frac{e^{-k\left(t \right) \left|x \right|}}{a^2 \left(t \right) k \left(t \right)}   
\\
\times
\left[\left|x\right| \left( a \left(t \right)^2 k\left(t \right)^2 +  {\rm sech^2} \left[ \frac{\left|x \right|}{a \left(t \right)}\right] \right) \right.
\left.+a\left(t \right)\left[ 2 + \left|x \right| k\left(t \right) \right]{\rm tanh}\left[\frac{\left|x \right|}{a \left(t \right)} \right]\right],
\end{multline}
where we have substituted~(\ref{eq:wavevec}) and~(\ref{eq:widtha}). This is a strictly positive quantity, representing a tendency of flow to enhance mode growth. An example eigenfunction evolution is shown in Figure~\ref{fig:psi1}, which shows that the eigenfunction tends to grow as time increases.  Note that this growth is clearly non-separable; for example, the eigenfunction's value does not change at $x=0$ while it does change elsewhere.  Thus, this effect cannot be expressed as a correction to the separable exponential growth rate, as in~(\ref{eq:previous}).

The reader may also be interested in the nature of higher order corrections of the flow given by the equations at $\mathcal{O} \left(\epsilon\right)$.  At this order, we have
\begin{multline}
\label{eq:higherorder}
\gamma_0 \left[ \partial_\tau \Psi_{1,0}\left(x,\tau\right)-x \partial_x \Psi_{1,0} \left(x,\tau\right) +  \gamma \left( \tau \right) \Psi_{1,1} \left( x,\tau \right)  \right]
  \\
 = \partial_x \psi_0\left(x,\tau \right) i k\left(\tau \right) \Phi_{1,1} \left(x,\tau\right) + \eta \left[ \partial_x^2 -k ^2\left(\tau \right)\right] \Psi_{1,1} \left(x,\tau\right)
\end{multline}
and
\begin{multline}
\label{eq:higherorder2}
\gamma_0 \left(\left[\partial_x^2 - k^2\left(\tau\right) \right]\partial_\tau \Phi_{1,0} \left(x,\tau\right) \right.
\left.
- x \partial_x \left[ \partial_x^2 -k^2 \left( \tau \right)\right] \Phi_{1,0} \left(x,\tau \right) + 2 k^2\left(\tau \right) \Phi_{1,0} \left(x,\tau \right)\right.
 \\
\left. + \gamma\left( \tau \right) \left[\partial_x^2 - k^2\left(\tau\right) \right] \Phi_{1,1} \left(x,\tau\right) \right)
 \\
= \partial_x \psi_0\left(x,\tau \right) i k\left(\tau \right) \left[\partial_x^2 - k^2 \left(\tau\right) \right] \Psi_{1,1} \left(x,\tau\right)
 -\partial_x^3 \psi_0 \left(x,\tau \right)i k\left(\tau \right) \Psi_{1,1} \left(x,\tau\right).
\end{multline}
Since $\Psi_{1,0}$, $\Phi_{1,0}$, and $\gamma \left( \tau \right)$ are found from the $\mathcal{O} \left( \epsilon^0 \right)$ equations, these expressions are coupled differential equations in $x$ only for $\Psi_{1,1}$ and $\Phi_{1,1}$ which may be solved to find small, non-separable corrections to the tearing mode eigenfunctions. Note that even these perturbations grow exponentially with the tearing growth rates given by~(\ref{eq:modFKR}),~(\ref{eq:modcop}), and~(\ref{eq:modmax}), as appropriate.

Thus, we have shown that in a developing current sheet in the regime where $\gamma \gg \Gamma_0$, flows affect the separable exponential tearing growth only implicitly, by making $a$, $B_y$, and $k$ time-dependent, and not through a correction  that explicitly depends on flow of the type represented in Eq.~(\ref{eq:previous}).  The flows will add slow, non-separable time dependence to the tearing mode eigenfunctions, and we have stated the lowest-order strength of this effect in Eqs.~(\ref{eq:changeeig}) and~(\ref{eq:changeeig2}); additional corrections could in principle be calculated from Eqs.~(\ref{eq:higherorder}) and~(\ref{eq:higherorder2}). But this slow time dependence is overwhelmed by the strong, exponential tearing mode growth rate.

\section{Nonlinear Theory}
\label{sec:nonlinear}
As the linear stage develops, it will eventually reach a point where the width of the forming magnetic island is comparable to the width of the inner resistive layer and the linear approximation is no longer valid. A Coppi mode will at this point enter Waelbroeck X-point collapse (which will not be affected by the flow because it proceeds at a rate comparable to that of the fastest tearing mode, much faster than the flow rate~\citep{waelbroeck1993onset,loureiro2005x}). An FKR mode will enter the Rutherford phase \citep{rutherford1973nonlinear}, which we examine here. 

We adapt the original flow-free Rutherford theory to the evolving system using the coordinate transformation~(\ref{eq:transformation}) and look for the growth of the tearing mode island width~$W$.  This process reveals two effects that modify the flow-free theory in the evolving case.  First, adding time dependence to  $k$ and $a$ leads to an increase in the island width's Rutherford growth rate through time. However, proper inclusion of the changing background and flow-dependent terms also yields a narrowing of the island width due to the compression of the equilibrium. This effect does not change the functional dependence of the growth but reduces it slightly by a constant scalar factor.

In the Rutherford phase, nonlinear instability behavior is analyzed through Ohm's law~(\ref{eq:ohm}). In flow-free analysis, the terms $-B_{0y} ik \phi_1$ and $ -\eta k^2 \psi_1$ are considered negligible in~(\ref{eq:ohm}); the addition of flow will not change this, and so we can state the perturbation's governing equation as 
\begin{equation}
\label{eq:grow}
\partial_t \psi_1 - \Gamma_0 \left(t \right) x \partial_x \psi_1 = \eta \partial_x^2 \psi_1.
\end{equation}
Taking the case of constant $\Gamma_0$ (similar results hold for slowly-varying $\Gamma_0$ and the techniques in this section may be adapted for non-slowly-varying $\Gamma_0$ as well) and transforming coordinates in this equation according to~(\ref{eq:transformation}) yields 
\begin{equation}
\label{eq:ruth}
\partial_{t'} \psi_1= \eta e^{2\Gamma_0 t'}\partial_{x'}^2 \psi_1.
\end{equation}
In this coordinate system, the perturbation growth is caused by a resistivity that appears to grow with time.  This apparent growth is due to the sharpening of gradients, caused by the compressing flow. We thus apply the flow-free procedure~\citep{rutherford1973nonlinear,biskamp2005magnetic} in these coordinates, adapting it to deal with the time dependence.  

To do this, we will need to determine the behavior of island width $W$ and tearing mode stability index $\Delta'$ in both the unprimed and primed systems.
Recall the expression for the tearing mode island width in the unprimed system:
\begin{equation}
W  = 4 \sqrt{\frac{-\psi_1}{\partial_{x}^2 \psi_0}}.
\end{equation}
Note that since $\partial_x^2 \psi_0$ increases with time, the island width of a perturbation with constant $\psi_1$ decreases with time. In the primed frame, the island width $W'$ is defined in terms of the constant-in-time $\partial_{x'} \psi_0$, so that 
\begin{equation}
\label{eq:widthdef}
W'  = 4 \sqrt{\frac{-\psi_1}{\partial_{x'}^2 \psi_0}} = W e^{\Gamma_0 t},
\end{equation}
which we could also see by considering that lengths in the unprimed frame are related to lengths in the primed frame by $x' = xe^{\Gamma_0 t}$.
For the tearing mode stability index $\Delta'$, recall the definition \citep{furth1963finite} $\Delta' \equiv \lim_{\epsilon \rightarrow 0} \left[\partial_x \psi_1\left( \epsilon \right) - \partial_x \psi_1 \left(-\epsilon \right)\right]/\psi_1$.  Since this quantity is an inverse length, it transforms according to
\begin{equation}
\left( \Delta ' \right)' = e^{-\Gamma_0 t} \Delta '.
\end{equation}
For the Harris sheet~(\ref{eq:harris}) for $k\left(t \right) a \left(t \right) \ll 1$, we can then evaluate
\begin{equation}
\left( \Delta ' \right)' =e^{-\Gamma_0 t} \frac{1}{k \left(t \right) a \left(t \right)^2} = \frac{e^{2\Gamma_0 t}}{k_0 a_0^2}.
\end{equation}

Returning to equation~(\ref{eq:ruth}), we can rewrite the second derivative of the flux function in terms of $W'$ and $\left( \Delta ' \right)'$, so that 
\begin{equation}
\partial_{t'} \psi_1 \sim \frac{\eta e^{2\Gamma_0 t' } \left(\Delta ' \right)' \psi_1}{W'} = \frac{\eta e^{4\Gamma_0 t'  } \psi_1}{k_0 a_0^2 W'}.
\end{equation} 
Relating $W'$ to $\psi_1$ using~(\ref{eq:widthdef}) (noting that $\partial_{x'}^2 \psi_0$ is time-independent) then gives 
\begin{equation}
\partial_{t'} W' \sim \frac{ \eta e^{4 \Gamma_0 t'} }{k_0 a_0^2}.
\end{equation}
Using~(\ref{eq:widthdef}) to also state our result in terms of island width in the unprimed system yields
\begin{equation}
\label{eq:Wev}
\partial_t W \sim \eta \Delta'  - \Gamma_0 W,
\end{equation}
where we have  stated our result in terms of $\Delta'\left(t\right) \sim 1/\left[ k\left(t \right) a \left( t \right)^2\right]\sim e^{3\Gamma_0 t}/ \left(k_0 a_0^2 \right)$.

This reveals a significant acceleration in the growth of the island width due to the change with time of the tearing mode stability index.  This effect was used in the treatment of the nonlinear phase in~\citep{PhysRevLett.116.105003} through its  inclusion of time dependence on $L$ (or equivalently $N/k$) and $a$ in nonlinear growth rates.  However, the treatment in~\citep{PhysRevLett.116.105003} did not consider that  the apparent width of the island is also decreased due to the flow and the changing equilibrium, which is reflected in our subtraction of $\Gamma_0 W$.  To further understand the effect of this term, we can first integrate~(\ref{eq:Wev}) to find
\begin{equation}
\label{eq:intW}
W\left(t\right) \sim \left(W_0 - \frac{\eta}{4 a_0^2 \Gamma_0 k_0} \right)e^{-\Gamma_0 t} + \frac{\eta e^{3\Gamma_0 t}}{4 a_0^2 \Gamma_0 k_0},
\end{equation}
with $W_0$ the island width at the start of the nonlinear phase, here defined as $t=0$.  The equivalent integration for~(\ref{eq:Wev}) without the subtracted $\Gamma_0 W$ term gives
\begin{equation}
\label{eq:intW2}
W\left(t\right) \sim W_0 - \frac{\eta}{3 a_0^2 \Gamma_0 k_0} + \frac{\eta e^{3\Gamma_0 t}}{3 a_0^2 \Gamma_0 k_0}.
\end{equation}
For large times, the behavior of both~(\ref{eq:intW}) and~(\ref{eq:intW2}) is dominated by their final terms, which differ only by a factor of $3/4$. The functional behavior of the nonlinear stage is thus the same in the evolving case, and using the flow-free equations with time dependence added, as done in~\citet{PhysRevLett.116.105003}, will only overestimate the growth by a scalar factor of order unity.

\section{Conclusion: Effect of flow in the context of current sheet development}
\label{sec:story}
We have presented a self-consistent RMHD model for a current sheet forming due to flow and examined the development of the tearing instability in it. Here, we review our analysis. A current sheet configuration will begin its evolution in a state where its characteristic width $a$ is such that the gradients in the current sheet are not sufficient to cause strong tearing. In this regime, the behavior of any pre-existing magnetic islands in the sheet can be studied using ideal MHD. This study, conducted in Section~\S\ref{sec:ideal}, shows that the islands will be compressed in $x$ and stretched in $y$, with their aspect ratio increasing, by the flow. 

As the current sheet compresses, it will become more unstable to tearing perturbations. 
Tearing behavior becomes important roughly at the time $t \sim t_{cr}$ when the magnitudes of the rates controlling current sheet evolution and tearing are comparable, i.e. $\Gamma_0 \left( t_{\rm cr} \right)= \gamma_{0, \rm max}\left(t_{\rm cr} \right)$ \citep{PhysRevLett.116.105003}.  After this time, the multiple scale formulation of Section~\S\ref{sec:linear} will become progressively better justified as the continually narrowing current sheet causes the drive for the tearing instability to overwhelm the flow rate.  In this regime, the tearing mode may be approximately described by the flow-free growth rates and eigenfunctions, with time dependence trivially added to $k$, $B_y$, and $a$. Small corrections to the eigenfunctions may be computed, but even these will grow with the flow-free tearing mode growth rate. 

Neither of these two approaches is valid during the period of time immediately surrounding $t_{cr}$, when $\Gamma_0 $ and  $\gamma_{0, \rm max}$ are of roughly the same magnitude, so we cannot make any definitive conclusions about the effects of flows during this time.\footnote{We note that for Alfv\'{e}nic driving flows $\Gamma_0 \sim v_{A}/L$, this period of time occurs when the sheet is at the critical aspect ratio identified by~\citet{pucci}.}   However, we can estimate the error that this uncertainty introduces. Namely, during this period the islands grow with a tearing growth rate $\gamma_{0, \rm max}\left(t\right) \sim \Gamma_0\left(t_{cr}\right)$.  But the current sheet equilibrium, and hence the tearing growth rate, whatever its form, evolve on the time scale $\Gamma_0^{-1}$.  Thus, the duration $\Delta t$ of the interval during which the two rates themselves are comparable is also of order $\Gamma_0^{-1}$. The factor by which the islands grow during this interval, given by
\begin{equation}
\exp{ \left[ \int \limits_{t_{\rm cr} - \Delta t/2}^{t_{\rm cr} + \Delta t/2}\gamma_{0,\rm max} \left(t \right) dt \right]},
 \end{equation}
can then be estimated as $e$ to a power of order unity.  This is a finite factor, independent of resistivity, which will not have a drastic effect on island development, and thus the ideal and linear treatments together cover the most significant periods of island growth.

At the end of the linear stage, when the width of the magnetic islands exceeds that of the inner resistive layer, the mode will enter the nonlinear regime, described in Section~\S\ref{sec:nonlinear}. Here, a simple coordinate transformation shows that to get a growth rate in the evolving sheet, one must both add time dependence to $k$ and $a$ in the flow-free expressions and subtract a term due to the strengthening of the background magnetic field. This will slightly slow the nonlinear growth relative to expressions studied in \citet{PhysRevLett.116.105003}.

Overall, this paper provides a detailed framework for understanding how the tearing mode behaves in an evolving current sheet.  While this analysis identifies some subtle effects not considered in \citet{PhysRevLett.116.105003}, like the subtracted term in the nonlinear stage and the small corrections to the linear eigenfunctions, all in all it strengthens confidence in the use of tearing mode scalings developed in a stationary configuration to study the tearing mode in an evolving sheet, and thus in the results of several recent studies of forming current sheets \citep{pucci,PhysRevLett.116.105003,comisso2016general,comisso2017plasmoid,huang2017plasmoid}. 

Future work based on this paper might consider two particular areas. First, our results are strictly limited to plasmas in which uniform-resistivity RMHD is valid.  This excludes collisionless, and semi-collisionless, plasmas; that is, plasmas for which the resistive layer width of the current sheet is smaller than the ion skin depth $d_i \equiv c /\omega_{pi}$, with $\omega_{pi}$ the ion plasma frequency, or the ion sound gyroradius $\rho_s \equiv \sqrt{\left(T_i +T_e\right)m_i}/\left(q_i B_T\right)$, where $T_i$ and $T_e$ are the ion and electron temperatures, respectively, $m_i$ is the ion mass, $q_i$ is the ion charge, and $B_T$ is the total magnetic field (including the guide field in the $\hat{z}$ direction).  For the fastest-growing linear mode in the sheet, found at the intersection of the FKR and Coppi regimes \citep{furth1963finite,coppi1976resistive}, the condition on the validity of our RMHD treatment may be stated in terms of Lundquist number:
\begin{equation}
\label{eq:col}
S_a \ll 
\begin{cases} \left( \frac{a}{d_i} \right)^4  & \text{if } \rho_s < d_i , \\ 
 \left(\frac{a}{\rho_s} \right)^4 & \text{otherwise}.
\end{cases}
\end{equation} 
Extending our work to collisionless, or semi-collisionless, regimes where~(\ref{eq:col}) does not hold would require a distinct analysis from that in this paper, though certain techniques that we use could likely be adapted. Such an extension is important given the weak collisionality of many plasmas in which reconnection occurs, including the magnetosphere \citep{ji2011phase}.

Second, we have restricted the analysis in this paper to current sheets with straight magnetic fields due to the challenges present in formulating a two dimensional current sheet model and in Fourier analyzing perturbations in a two dimensional equilibrium.  Since we study a forming current sheet where reconnection is just beginning, $\hat{x}$ magnetic components resulting from reconnected field do not yet exist.  However, it is possible that the background magnetic field from which the sheet forms is curved; indeed this is the case in the Earth's magnetotail \citep{petrukovich2015current}, and in astrophysical situations where current sheet formation occurs because of field-line opening due to differential rotation (see, e.g., \citet{uzdensky2004magnetic}  and references therein). An extension of our work could consider such curved fields, perhaps by employing computational methods.  

Finally, combining these two areas by studying current sheets with curved magnetic fields in collisionless plasmas is also of significant interest, as previous work has found that inclusion of two dimensional effects leads to new behavior in collisionless reconnection \citep{lembege1982stability,sitnov2011onset}.

The authors acknowledge support from the National Science
Foundation Graduate Research Fellowship under Grant No. DGE-1122374; the NSF-DOE Partnership in Basic Plasma
Science and Engineering, Award No. DE-SC0016215; DOE Award No. DE-SC0008409; and NASA Award No. NNX16AB28G. D.A.U.  gratefully acknowledges the hospitality of the Institute for Advanced Study and the support from the Ambrose Monell Foundation.

\bibliographystyle{jpp}
\bibliography{bib3}
\end{document}